\documentstyle[12pt]{article}
\begin{document}


\renewcommand{\textfraction}{0.2}    
\renewcommand{\topfraction}{0.8}

\renewcommand{\bottomfraction}{0.4}
\renewcommand{\floatpagefraction}{0.8}
\newcommand\mysection{\setcounter{equation}{0}\section}

\def\baeq{\begin{appeq}}     \def\eaeq{\end{appeq}}
\def\baeeq{\begin{appeeq}}   \def\eaeeq{\end{appeeq}}
\newenvironment{appeq}{\beq}{\eeq}
\newenvironment{appeeq}{\beeq}{\eeeq}
\def\bAPP#1#2{
 \markright{APPENDIX #1}
 \addcontentsline{toc}{section}{Appendix #1: #2}
 \medskip
 \medskip
 \begin{center}      {\bf\LARGE Appendix #1 :}{\quad\Large\bf #2}
\end{center}
 \renewcommand{\thesection}{#1.\arabic{section}}
\setcounter{equation}{0}
        \renewcommand{\thehran}{#1.\arabic{hran}}
\renewenvironment{appeq}
  {  \renewcommand{\theequation}{#1.\arabic{equation}}
     \beq
  }{\eeq}
\renewenvironment{appeeq}
  {  \renewcommand{\theequation}{#1.\arabic{equation}}
     \beeq
  }{\eeeq}
\nopagebreak \noindent}

\def\eAPP{\renewcommand{\thehran}{\thesection.\arabic{hran}}}

\renewcommand{\theequation}{\arabic{equation}}
\newcounter{hran}
\renewcommand{\thehran}{\thesection.\arabic{hran}}

\def\bmini{\setcounter{hran}{\value{equation}}
\refstepcounter{hran}\setcounter{equation}{0}
\renewcommand{\theequation}{\thehran\alph{equation}}\begin{eqnarray}}
\def\bminiG#1{\setcounter{hran}{\value{equation}}
\refstepcounter{hran}\setcounter{equation}{-1}
\renewcommand{\theequation}{\thehran\alph{equation}}
\refstepcounter{equation}\label{#1}\begin{eqnarray}}

\newskip\humongous \humongous=0pt plus 1000pt minus 1000pt
\def\caja{\mathsurround=0pt}


\title{
       {\Large
                 {\bf
 EMC effect with different oscillator-model
parameters \textit{$h\omega$}\ for different shells by considering
difference between proton and neutron structure functions.
                 }
         }
      }

\author{\vspace{1cm}\\
{\small F. Zolfagharpour\thanks {e-mail:
zolfagharpour@uma.ac.ir.}
} \\
{\small  Department of physics, University of Mohaghegh Ardabili
, P.O. Box 179, Ardabil, Iran}}
\date{}
\begin{titlepage}
\maketitle

\begin{abstract}
The magnitude of binding energy used in the conventional nuclear
theory to explain the EMC experimental data, seems to be larger
than the one expected. In this paper to get sufficient depletion
in the binding energy, different oscillator-model parameters $
\textit{$h\omega$}\ $ for different shells and the proton
(neutron) structure function that have good agrement with
experimental data are used. The extracted results for
$^4He,~^{12}C,~^{40}Ca$ and $^{56}Fe$ nuclei show that one can get
improved results in medium x ranges by less binding energy.
\end{abstract}

~~~PACS: 13.60.Hb, 21.10.Pc, 25.30.-c

~~~Keywords: EMC effect, Nuclear structure function, Shell model
\end{titlepage}

\section{Introduction}

In 1983 the European Muon Collaboration (EMC) reported \cite{1}
their measurement of muon scattering cross sections per nucleon of
iron to deuterium nuclei. The ratios were clearly different from
unitary and totally unexpected because of high momentum transfer
involved ($Q^2$ up to 200 GeV'). Several phenomena were proposed
to explain theoretically the EMC effect and each of them had some
successes in limited x (the Bjorken variable) range. The
conventional nuclear theory proposed by Akulinichev et al,
\cite{2} and Dunne et al, \cite{3} seems to be able to explain the
experimental data in medium and large x ranges. Akulinichev et al,
used a harmonic-oscillator model with the energy levels taken from
the compilation of experimental data for the (e, e'p) reaction
\cite{4}. They showed that the nuclear interaction of nucleons
resulting in their binding and Fermi motion, plays an important
role in deep-inelastic lepton-nuclear scattering. But the average
value of the potential for these oscillator levels was about -60
MeV. Dunne and Thomas also used a harmonic oscillator model with
the levels fitted to measured separation energies \cite{3}. So
they failed to get sufficient depletion for corresponding single
particle energy. If in that time someone had used measured
separation energies within the naive single-particle model, the
discrepancy would have gotten worse. The authors of references
\cite{2,3}, used only one oscillator-model parameter
$\textit{$h\omega$}$ for different shells inside the investigated
nuclei. So they tried to understand the EMC effect in term of the
change in the mass scale of a bounded nucleon \cite{44}. Also the
extracted results by the light front formalism that developed by
Miler and Smith \cite{44b,44c}, is not consistent with the binding
effect not only in the magnitude of the effect, but also in the
dependence on the number of nucleons. It should be mentioned that
they used not only one momentum distribution for any nuclear level
but also one nucleon structure function parametrization. The used
parametrization for nucleon structure function in reference
\cite{44c} considers no difference between protons and neutrons
structure function inside the investigated nuclei. To compare it
with other parametrization of nucleon structure functions, it is
plotted in figure 1 by dash curve. Also the convolution approach
represented in reference \cite{44d}, no difference between protons
or neutrons momentum distribution on different state has been
considered. As different shells have different root mean square
radius one can choose different oscillator-model parameters $
\textit{$h\omega$}$ for different shells inside a nucleus. Also it
should be mentioned that the nucleon structure function used by
the authors of references \cite{2,3,44b,44c}, is far from the
experimental data (see figure 1). So in this paper it is tried to
extract the EMC results in the conventional nuclear theory for the
nuclei $^{56}$Fe, $^{40}$Ca, $^{12}$C, $^{4}$He, first by
considering `the different oscillator-model parameters $
\textit{$h\omega$}$ for different shells related to their root
mean square radius', and second: `the free GRV's \cite{5} neutron
and proton structure functions that have good agreement with
experimental data'.

\section{Theoretical formalism}
One can obtain the nuclear structure function of nucleus by main
equation \cite{2}
\begin{equation}
F^{A}_{2}(x)=\sum_{N=\textrm{n},\textrm{p}}\sum_{nl}\hspace{1mm}\int_{x}^{\infty}dz
{g_{nl}^{N}}\textit{f}\hspace{1mm}^{N}(z)_{nl} F^{N}_{2}(x/z),
\end{equation}
where the first sum is over the proton and neutron cases. The
second sum is over the quantum number of each energy levels. The
$g_{nl}^{N}$ is the occupation number of energy level
$\epsilon_{nl}$ for proton (N=\textrm{p}) and neutron
(N=\textrm{n}). The nucleon distribution inside the nucleus define
as
\begin{equation}
 \textit{f}\hspace{1mm}^{N}(z)_{nl}=
 \int_{|m_{_N} (z-1)-\epsilon_{nl}|}^{\infty}dp\hspace{1mm}p
 m_{_N}|\phi_{nl}(p)|^2/(2\pi)^2,
\end{equation}
with $z=p_{nl}q/m_{_N}q_0$ and $x=Q^2/2m_{_N}q_0$ the bjorken
variable for free nucleon. The effects of the momentum and energy
distribution of the nucleon in the nucleus are included in Eq. (2)
through $\phi_{nl}(p)$ and $\epsilon_{nl}$, respectively. The
magnitude of  nuclear binding energy ($\epsilon_{nl}$) mainly
effects the structure functions in the intermediate x region. The
function $\textit{f}\hspace{1mm}^A(z)_{nl}$ describes the momentum
and energy distribution of nucleons inside nuclei and satisfies
the normalization rule
\begin{equation}
\sum_{N=\textrm{n},\textrm{p}}\sum_{nl}\hspace{1mm}\int_{0}^{\infty}dz
{g_{nl}^{N}}\textit{f}\hspace{1mm}^{N}(z)_{nl} =A.
\end{equation}
Akulinichev et al, \cite{2} used harmonic-oscillator nuclear  wave
function to calculated $\textit{f}\hspace{1mm}^A(z)_{nl}$. For the
oscillator-model parameter $\hbar\omega$ they used only one
parameter for any quantum number \textit{n}, \textit{l}. We knew
in the heavier nuclei the deeper closed shells have different root
mean square radius and for this purpose, in the
harmonic-oscillator model one could use \cite{6} (see appendix A)
\begin{equation}
<r^2>_{nl}\hspace{1mm}=\frac {1}{\alpha^2}(2n+l+\frac{3}{2}),
\end{equation}
where
\begin{equation}
\alpha^2=\frac {m_{_N}\omega}{\hbar}
\end{equation}
and by considering $m_{_N}=938.905$ $MeV$, one can find in the
natural unit
\begin{equation}
{\hbar\omega}=\frac {41.33}{<r^2>_{nl} }(2n+l+\frac{3}{2}),
\end{equation}
where ${<r^2>}^{1/2}_{nl}$ and ${\hbar\omega}$ expressed
respectively in Fermi and MeV unit. Table 1 shows the calculated
$\hbar\omega$ for the nuclei that are investigated here and the
table 2 contains the brackets that shows the occupation number for
different levels.

The resulting expression for the nucleon distribution
$\textit{f}\hspace{1mm}^{N}(z)_{nl}$ inside the nucleus is \cite{2}

\begin{eqnarray}
&& \textit{f}\hspace{1mm}^{N}(z)_{nl}=\frac {1}{2}\bigg (\frac
 {\hspace{1mm}m_{_N}}{\hbar\omega}\bigg)^{1/2}\frac
 {n!}{\Gamma(n+l+\frac{3}{2})}\sum^{n}_{t_1=0}\sum^{n}_{t_2=0}
 \frac{(-1)^{t_1+t_2}}{t_1!t_2!}
 \Big{(}^{n+l+\frac{1}{2}}_{\hspace{2mm}n-t_1} \Big{)}
 \nonumber \\*
&& \hspace{18mm} \times
\Big{(}^{n+l+\frac{1}{2}}_{\hspace{2mm}n-t_2} \Big{)} \Gamma
\large{[}l+t_1+t_2+1,\frac{m_{_N}}{\hbar\omega}
 \large{(}z-1-\frac{\epsilon_{nl}}{m_{_N}}\large{)}^2\large{]}.
\end{eqnarray}

For the $ F^{N}_{2}(x/z)$ the GRV$^{^,}$s LO free proton and
neutron structure functions parameterizations used \cite{5}
\begin{eqnarray}
&& \frac
 {\hspace{1mm}1}{x} F^{ep}_{2}(x,Q^2)= \sum_q e^2_q \bigg\{q(x,Q^2)+\overline{q}(x,Q^2)+\frac{\alpha_s(Q^2)}{2\pi}
  \nonumber \\*
&& \hspace{28mm} \times
\bigg[C_{q,2}\hspace{1mm}^*(q+\overline{q})+2C_{g,2}\hspace{1mm}^*g\bigg]
\bigg\}\nonumber\\* &&  \hspace{28mm}
+\frac{1}{x}F^c_2(x,Q^2,m^2_c)
\end{eqnarray}
with
\begin{eqnarray}
&& C_{q,\hspace{1mm}2}(z)=\frac
 {4}{3}
 \bigg[\frac{1+z^2}{1-z}\bigg(\ln\frac{1-z}{z}-\frac{3}{4}\bigg)+\frac{1}{4}(9+5z)\bigg]_{+},
  \nonumber \\*
\end{eqnarray}
\begin{eqnarray}
&& C_{g,\hspace{1mm}2}(z)=\frac
 {1}{2}
 \bigg[(z^2+(1-z)^2)\ln\frac{1-z}{z}-1+8z(1-z)\bigg]
  \nonumber \\*
\end{eqnarray}
and the $\sum_{q}\frac{}{}$ extended over all light quarks
$q=u,d,s$. The convolution product and the convolution product
with $[\hspace{3mm}]_+$ are defined as usual
\begin{eqnarray}
&& C^*q=\int_x^1\frac{dy}{y}C\bigg(\frac{x}{y}\bigg)q(y,Q^2),
   \nonumber \\*
\end{eqnarray}
\begin{eqnarray}
&&
\int_x^1\frac{dy}{y}f\bigg(\frac{x}{y}\bigg)_+g(y)=\int_x^1\frac{dy}{y}f\bigg(\frac{x}{y}\bigg)
\bigg[g(y)-\frac{x}{y}g(x)\bigg]-g(x)\int_0^xdyf(y)
   \nonumber \\*
\end{eqnarray}

and
\begin{eqnarray}
&&
\frac{\alpha_s(Q^2)}{2\pi}=\frac{2}{\beta_0\ln(Q^2/\Lambda^2)}-\frac{2\beta_1}{\beta_0^3}
\frac{\ln\ln(Q^2/\Lambda^2)}{[\ln(Q^2/\Lambda^2)]^2}
   \nonumber \\*
\end{eqnarray}

with $\beta_0=11-2\textit{f}/3$, $\beta_1=102-38\textit{f}/3$ and
$\Lambda^{^{{\scriptsize\textit{f}}\hspace{1mm}=4}}_{{\scriptsize
LO}}=0.2$ GeV. The LO charm quark contribution in Eq. (8) is
defined as

\begin{eqnarray}
&&\frac{1}{x}F^c_2(x,Q^2,m^2_c)=2e^2\frac{\alpha_s({\mu'}
^2)}{2\pi}\int_{ax}^1\frac{dy}{y}C^2_{g,\hspace{1mm}2}\bigg(\frac{x}{y},\frac{m_c^2}{Q^2}\bigg)g(y,{\mu'}^2)
   \nonumber \\*
\end{eqnarray}
with $a=1+4m_c^2/Q^2$ and in LO
\begin{eqnarray}
&& C^c_{g,\hspace{1mm}2}(z,\frac{m_c^2}{Q^2})=\frac
 {1}{2}
 \bigg\{\bigg[z^2+(1-z)^2)+z(1-3z)\frac{4m_c^2}{Q^2}-z^2\frac{8m_c^4}{Q^4}\bigg]
  \nonumber \\*
 &&
 \hspace{28mm}\ln\frac{1+\beta}{1-\beta}+\beta\bigg[-1+8z(1-z)-z(1-z)\frac{4m_c^2}{Q^2}\bigg]\bigg\},
\end{eqnarray}
where $\beta^2=1-(
4m_c^2/{Q^2})z(1-z)^{-1}$,$\hspace{2mm}\mu'=4m_c^2$
\hspace{1mm}and $m_c=1.5$ GeV. \hspace{1mm}Figure 1 shows the
extracted GRV$^{^,}$s free proton and neutron structure function
that used in Eq. (1). The related LO GRV$^{^,}$s parton
distribution Fortran code can be found in \cite{7}. The
calculated ratio
\begin{eqnarray}
R^A_{_{EMC}}(x)=F^A_{2}(x)_{Per\hspace{1mm}
nucleon}/F^{^2H}_2(x)_{{Per\hspace{1mm}nucleon }},
\end{eqnarray}
presented in figure 2. To better express deformation of bonded
nucleon structure function in comparison with free nucleon
structure function, the ratio
\begin{eqnarray}
R^A(x)=\frac{F^A_{2}(x)}{ZF_2^{\textrm{p}}+NF_2^{\textrm{n}}}
\end{eqnarray}

is calculated. Where here N is the number of neutron and the Z is
the atomic number so $A=Z+N$. The extracted results for $R^A(x)$
are plotted in figure 4.

\section{result, discussion and conclusion}
Figure 1 shows the free nucleon structure function (read curve)
that is used by S.~V. Akulinichev et al, \cite{2}. This is far
from the GRV$^{^,}$s free nucleon structure function
($(F^\textrm{p}_{2\hspace{1mm},GRV}+F^\textrm{n}_{2\hspace{1mm},GRV})/2$)
that is shown by brown curve [5,7]. It seems, if the corrected
nucleon structure functions were used in Eq. (1), the calculated
$R^A_{EMC}(x)$ could be improved. It is necessary to notice that
we can not use the nucleon structure function like read curve even
brown in figure 1, instead of proton and neutron structure
functions for the nuclei like nucleus $^{56}$Fe because of
Z$\neq$N. In this paper for extracting all the results, the
GRV$^{^,}$s proton and neutron structure functions are used. The
difference between the root mean square of shells encourages me to
consider different $\hbar\omega$ parameters for different shells
inside a nucleus (see table 1). The plotted results in figure 2,
3, 4 extracted by using: a) Proton and neutron structure functions
that have good agreement with experimental data, b) The
$\hbar\omega$ parameter that is related to shell$^{^,}$s root mean
square for investigated nuclei. By considering the cases mentioned
above, the binding energies $\varepsilon_{nl}$ (see table 2) that
are used to obtain the results in figure 2, seems to be less than
those used in previous published paper. For example, compare
$\varepsilon_{nl}=-40\hspace{1mm}MeV$ for $^{56}Fe$ \cite{2,3,11}
with $\varepsilon_{nl}=-26\sim-32\hspace{1mm}MeV$ which are used
here to extract the result for the $^{56}$Fe that is shown in
figure 2 (see table 2). The figure 3 shows that with this new used
$\hbar\omega$ parameters (see table 1), the contribution of only
Fermi motion effects specially for $^{56}Fe$ (full curve) is
similar to the previous work \cite{10} that is plotted by dash
curves in this figure but we should notice the full curves rise up
about $0.05\sim0.02$ more than related Bjorken parameter scale.
The reason of this effect is that the $\hbar\omega$ parameters
related to the root mean square radius are a bit larger than the
$\hbar\omega$ parameters for example used in 2. We see that the
used harmonic oscillator parameters by Akulinichev, et al., are
related to larger radius than those that the experimental data
shows. In the medium and large x ranges the neutron structure
function is smaller than the proton structure function so the
extra 4 neutrons in the $^{56}Fe$ nucleus cause the EMC ratio of
iron comes down about 3 percent more than the other nuclei even by
putting $\varepsilon_{nl}=0 $ in medium x range. In the figure 4
according to the Eq.\hspace{2mm}(17) the ratio of investigated
nuclei$^{^,}$s structure function to simple sum of equal free
protons and neutrons GRV$^{^,}$s structure function plotted. One
can find a bit difference in the extracted results for each
nucleus because of different radius in the case of neglecting
binding energy and considering only Fermi motion effect (dash
curve). We can trust the results from $x\sim0.2$ to $x\sim0.7$
\cite{12} But when the effect of binding energy added to the Fermi
motion effect the difference between the ratio for different
nuclei near the $x\sim0.7$ is decreased (full curves). The
magnitude of difference between full curve  near the $x\sim0.7$ is
small enough that one interpret this effect as a result of
saturation for large x ranges. The figure 4 shows the binding
effect plays important role in medium x ranges.

\section*{Appendix A}
To calculate root mean square radius we start from calculated
total wave function for harmonic oscillator potential
\begin{eqnarray}
\psi(r)=R(r)Y(\theta,\varphi) \nonumber\\
=\frac{u(r)}{r}Y(\theta,\varphi).
\end{eqnarray}
The equation governing the radial motion is
\begin{eqnarray}
-\frac{\hbar^2}{2m}\frac{d^2u(r)}{dr^2}+[l(l+1)+V(r)]u(r)=Eu(r).
\end{eqnarray}
in studing bound states, conditions have to be imposed on radial
solution u(r):
\begin{eqnarray}
&& \lim_{r \to \infty}u(r)\rightarrow0 \nonumber
\\* && u(0)=0.
\end{eqnarray}
the normalization of radial wave function leads to the integrals
\begin{eqnarray}
\int^\infty_0R^2(r)dr=\int^\infty_0u^2(r)dr=1.
\end{eqnarray}
for harmonic oscillator potential i.e., $V(r)=\frac{1}{2}mw^2r^2$
one can obtain the radial laguerre equation with the solution
\cite{13}
\begin{eqnarray}
u_{nl}(r)=N_{n,\hspace{1mm}l}\hspace{1mm}r^{l+1}e^{-\nu
r^2}L_n^{l+1/2}(2\nu r^2),
\end{eqnarray}
where $\nu=m\omega/2\hbar$ , $N_{n,\hspace{1mm}l}$ normalization
factor and $E=\hbar\omega(2n+l+\frac{3}{2})$. For root mean square
radius we have
\begin{eqnarray}
&& <r^2>_{nl}=\int \psi_{nl}^\ast(r) r^2 \psi_{nl}(r) dv \nonumber
\\* &&~~~~~~~~=\int\frac{u_{nl}(r)}{r}r^2\frac{u_{nl}(r)}{r}r^2
dr\int Y^\ast(\theta,\varphi)Y(\theta,\varphi)d\Omega \nonumber
\\* && ~~~~~~~~=\int^\infty_0 u_{nl}^2(r) r^2 dr \nonumber \\* &&
~~~~~~~~=\int^\infty_0 \bigg\{
N_{n,\hspace{1mm}l}\hspace{1mm}r^{l+1}e^{-\nu r^2}L_n^{l+1/2}(2\nu
r^2)\bigg\}^2r^2 dr \nonumber \\* &&
~~~~~~~~=\int^\infty_0N^2_{n,\hspace{1mm}l}\hspace{1mm}(r^2)^{l+1}e^{-2\nu
r^2}L_n^{l+1/2}(2\nu r^2)L_n^{l+1/2}(2\nu r^2)r^2dr.
\end{eqnarray}
If we put $x=2\nu r^2$ and
$dr=\frac{1}{4\nu}\sqrt{\frac{2\nu}{x}}dx$ then we have
\begin{eqnarray}
 &&<r^2>_{nl}=\int^\infty_0N^2_{n,\hspace{1mm}l}\hspace{1mm}(\frac{x}{2\nu})^{l+1}e^{-x}
 L_n^{l+1/2}(x)L_n^{l+1/2}(x)\frac{x}{2\nu}\frac{1}{4\nu}\sqrt{\frac{2\nu}{x}}dx \nonumber
\\* && ~~~~~~~~=N^2_{n,\hspace{1mm}l}\hspace{1mm}(\frac{1}{2\nu})^{l+1}\frac{\sqrt{2\nu}}{8\nu^2}\int^\infty_0
x^{l+\frac{3}{2}}e^{-x} L_n^{l+1/2}(x)L_n^{l+1/2}(x)dx,
\end{eqnarray}
where \cite{14}
\begin{eqnarray}
\int^\infty_0x^{k+1}e^{-x}
L_n^{k}(x)L_n^{k}(x)dx=\frac{(n+k)!}{n!}(2n+k+1).
\end{eqnarray}
So we can write
\begin{eqnarray}
 &&<r^2>_{nl}=N^2_{n,\hspace{1mm}l}\hspace{1mm}(\frac{1}{2\nu})^{l+1}\frac{1}{\sqrt{32\nu^3}}\int^\infty_0
x^{(l+\frac{1}{2})+1}e^{-x} L_n^{l+1/2}(x)L_n^{l+1/2}(x)dx.
\end{eqnarray}
By using Eq. (25) and putting $k=l+\frac{1}{2}$ , we have
\begin{eqnarray}
 &&<r^2>_{nl}=N^2_{n,\hspace{1mm}l}\hspace{1mm}(\frac{1}{2\nu})^{l+1}\frac{1}{\sqrt{32\nu^3}}
 \frac{(n+l+\frac{1}{2})!}{n!}(2n+l+\frac{3}{2}).
\end{eqnarray}
To calculate normalization factor
\begin{eqnarray}
 &&\int^\infty_0 u_{nl}^2(r) dr=1 \nonumber \\* &&
 ~~~~~~~~~~~~~~~~=\int^\infty_0\bigg\{
N_{n,\hspace{1mm}l}\hspace{1mm}r^{l+1}e^{-\nu r^2}L_n^{l+1/2}(2\nu
r^2)\bigg\}^2 dr \nonumber \\* &&
~~~~~~~~~~~~~~~~=N^2_{n,\hspace{1mm}l}\hspace{1mm}\int^\infty_0
(r^2)^{l+1}e^{-2\nu r^2}L_n^{l+1/2}(2\nu r^2)L_n^{l+1/2}(2\nu r^2)
dr \nonumber \\* &&
\end{eqnarray}
If we put $x=2\nu r^2$ then we have
\begin{eqnarray}
 &&
1=\int^\infty_0 u_{nl}^2(r) dr
=N^2_{n,\hspace{1mm}l}\hspace{1mm}\int^\infty_0
(\frac{x}{2\nu})^{l+1}e^{-x}L_n^{l+1/2}(x)L_n^{l+1/2}(x)\frac{1}{4\nu}\sqrt{\frac{2\nu}{x}}dx
\nonumber
\\* &&~ =N^2_{n,\hspace{1mm}l}\hspace{1mm}(\frac{1}{2\nu})^{l+1}\sqrt{\frac{1}{8\nu}}\int^\infty_0
x^{l+1}e^{-x}L_n^{l+1/2}(x)L_n^{l+1/2}(x)x^{-\frac{1}{2}}dx
\nonumber
\\* &&~ =N^2_{n,\hspace{1mm}l}\hspace{1mm}(\frac{1}{2\nu})^{l+1}\sqrt{\frac{1}{8\nu}}\int^\infty_0
x^{l+\frac{1}{2}}e^{-x}L_n^{l+1/2}(x)L_n^{l+1/2}(x)dx
\end{eqnarray}
where \cite{14}
\begin{eqnarray}
\int^\infty_0x^{k}e^{-x}
L_n^{k}(x)L_m^{k}(x)dx=\frac{(n+k)!}{n!}\delta_{n,\hspace{1mm}m}~,
\end{eqnarray}
so
\begin{eqnarray}
 &&
1=N^2_{n,\hspace{1mm}l}\hspace{1mm}(\frac{1}{2\nu})^{l+1}\sqrt{\frac{1}{8\nu}}\int^\infty_0
x^{l+\frac{1}{2}}e^{-x}L_n^{l+1/2}(x)L_n^{l+1/2}(x)dx\nonumber
\\*
&&~=N^2_{n,\hspace{1mm}l}\hspace{1mm}(\frac{1}{2\nu})^{l+1}\sqrt{\frac{1}{8\nu}}\frac{(n+l+\frac{1}{2})!}{n!}
\end{eqnarray}
and
\begin{eqnarray}
N^2_{n,\hspace{1mm}l}\hspace{1mm}=(2\nu)^{l+1}\sqrt{8\nu
}\frac{n!}{(n+l+\frac{1}{2})!}.
\end{eqnarray}
By substituting this normalization factor in Eq. (27) one can find
\begin{eqnarray}
 && <r^2>_{nl}=(2\nu)^{l+1}\sqrt{8\nu
}\frac{n!}{(n+l+\frac{1}{2})!}(\frac{1}{2\nu})^{l+1}\frac{1}{\sqrt{32\nu^3}}
 \frac{(n+l+\frac{1}{2})!}{n!}(2n+l+\frac{3}{2})\nonumber
\\* && ~~~~~~~~~~=\frac{1}{2\nu}(2n+l+\frac{3}{2})\nonumber
\\* && ~~~~~~~~~~=\frac{1}{2\frac{m\omega}{2\hbar}}(2n+l+\frac{3}{2})\nonumber
\\* && ~~~~~~~~~~=\frac{\hbar}{m\omega}(2n+l+\frac{3}{2})
\end{eqnarray}

\newpage

\section*{Figure captions}

Figure 1:  The LO GRV$^{^,}$s structure functions for proton
(blue) and neutron (pink) at $Q^{^2}=4\hspace{2mm}GeV^{^2}$
without charm quark contribution that is used in Eq. (1). Notice
to the difference between the GRV$^{^,}$s nucleon structure
function (brown) and the red curve that shows the used nucleon
structure function in refrence 2. The experimental data are taken
from \cite{8,9}.
\\ \\
Figure 2:  The extracted
$F^A_2(x)_{Per\hspace{1mm}nucleon}/F^{^2H}_2(x)_{Per\hspace{1mm}nucleon}$
for A=$^4He$, $^{12}C$, $^{40}Ca$ and $^{56}Fe$ from Eq. (1). The
$F^N_2(x)$ are taken  from GRV$^{^,}$ proton N=$\textrm{p}$ and
neutron N=$\textrm{n}$ structure functions according to the Eq.
(8) that plotted in figure 1. The used parameters $(g_{nl}^{\small
\textrm{p}},{g_{nl}^{\textrm{\textrm{n}}}},\epsilon_{nl})$ and the
calculated $\hbar\omega$ parameter are shown in table 1 and 2. The
experimental data without error bar are taken from \cite{7,10a}.
\\ \\
Figure 3:  The extracted $F^A_2(x)_{_{Per\hspace{1mm}
nucleon}}/\hspace{1mm}F^{^2H}_2(x)_{_{Per\hspace{1mm} nucleon}}$
for A=$^4He$, $^{12}C$, $^{40}Ca$ and $^{56}Fe$ from Eq. (1) by
considering only Fermi motion effect. The full curves were
obtained with the parameters explained in the caption of figure 2
by putting $\epsilon_{nl}=0$. The dash curves with the same color
for the same nucleus were obtained with the parameters
$\hbar\omega$ of Reference \cite{3}.
\\ \\
Figure 4:  The extracted $F^A_2/(NF^\textrm{p}+NF_2^\textrm{n})$
for A=4, 12, 40, and 56 from Eq. (1). The used parameters are the
same as parameters that are used to extract the results in figure
1. The dash curves with the same color for the same nucleus shows
the extracted results by considering only the Fermi motion effect
(i.e. $\varepsilon_{nl}=0$).
\newpage

\begin{table}
\renewcommand{\arraystretch}{1.5}
\addtolength{\arraycolsep}{-1pt}
$$
\begin{array}{c c c c c c c}
\hline \hline
Shell & ^2H &^4He& ^{12}C & ^{28}Si &^{40}Ca& ^{56}Fe \\
\hline
0s & (2.09,15.35)&(1.67,22.23)& (1.67,22.23)&(1.67,22.23)&(1.67,22.23)&(1.67,22.23) \\
0p &            &           & (2.44,17.36)&(2.44,17.36)&(2.44,17.36)&(2.44,17.36) \\
0d &            &           &            &(3.10,15.05)&(3.10,15.05)&(3.10,15.05) \\
1s &            &           &            &           &(3.48,11.95)&(3.48,11.95) \\
0f &            &           &            &           &           &(3.74,13.3) \\
\hline \hline
\end{array}
$$
\caption{The brackets contain
($<r^2>^{1/2}_{nl}$,\hspace{1mm}$\hbar\omega$). The
oscillator-model parameters \textit{$h\omega$} calculated from Eq.
(6). }
\renewcommand{\arraystretch}{1}
\addtolength{\arraycolsep}{-3pt}
\end{table}
\newpage
\begin{table}
\renewcommand{\arraystretch}{1.5}
\addtolength{\arraycolsep}{-0.5pt}
$$
\begin{array}{c c c c c c c}
\hline \hline
Shell & ^2H &^4He& ^{12}C & ^{28}Si &^{40}Ca& ^{56}Fe \\
\hline
0s & (1,1,-1)   &  (2,2,-15)& (2,2,-22)  &(2,2,-20)  &(2,2,-30)   &(2,2,-32) \\
0p &            &           & (4,4,-20)  &(6,6,-20)  &(6,6,-28)   &(6,6,-32) \\
0d &            &           &            &(6,6,-18)  &\hspace{4mm}(10,10,-26) &\hspace{4mm}(10,10,-30) \\
1s &            &           &            &           &(2,2,-25)   &(2,2,-28) \\
0f &            &           &            &           &            &\hspace{2mm}(6,10,-26) \\
\hline \hline
\end{array}
$$
\caption{The brackets contain $(g_{nl}^{\small
\textrm{p}},{g_{nl}^{\textrm{\textrm{n}}}},\epsilon_{nl}(MeV))$
for related shell. }
\renewcommand{\arraystretch}{1}
\addtolength{\arraycolsep}{-3pt}
\end{table}

\end{document}